\documentclass[%
 aip,
 amsmath,amssymb,
 reprint,%
]{revtex4-2}

\usepackage{graphicx}% Include figure files
\usepackage{dcolumn}% Align table columns on decimal point
\usepackage{bm}% bold math
\usepackage[utf8]{inputenc}
\usepackage[T1]{fontenc}
\usepackage{mathptmx}
\usepackage{etoolbox}
\usepackage{textcomp}
\usepackage{lineno}
\usepackage{ulem}
\usepackage{amssymb}
\usepackage[colorlinks,linkcolor=blue,
            citecolor=blue,
            hyperindex,
            pdfstartview=FitH,
            plainpages=false]
            {hyperref}

\newcommand{\BS}{Bi$_2$Se$_3$}
\newcommand{\YBCO}{YBa$_2$Cu$_3$O$_{7-\delta}$}
\newcommand{\Bi}{Bi$_2$Sr$_2$CaCu$_2$O$_{8+\delta}$}
\newcommand{\um}{\textmu m}

\makeatother
\begin{document}

%\preprint{AIP/123-QED}

\title{A sample-position-autocorrection system with precision better than 1 \um~in angle-resolved photoemission experiments}
% Force line breaks with \\
\author{Shaofeng Duan}
\author{Shichong Wang}
\author{Yuanyuan Yang}
\author{Chaozhi Huang}
\author{Lingxiao Gu}
\author{Haoran Liu}
\affiliation{Key Laboratory of Artificial Structures and Quantum Control (Ministry of Education),Shenyang National Laboratory for Materials Science, School of Physics and Astronomy, Shanghai Jiao Tong University, Shanghai 200240, China}
\author{Wentao Zhang}
\homepage{Author to whom correspondence should be addressed: wentaozhang@sjtu.edu.cn}
\affiliation{Key Laboratory of Artificial Structures and Quantum Control (Ministry of Education),Shenyang National Laboratory for Materials Science, School of Physics and Astronomy, Shanghai Jiao Tong University, Shanghai 200240, China}
\affiliation{Collaborative Innovation Center of Advanced Microstructures, Nanjing University, Nanjing 210093, China}

\date{\today}% It is always \today, today,

\begin{abstract}
We present the development of a high-precision sample-position-autocorrection system for photoemission experiments. A binocular vision method based on image pattern matching calculations was realized to track the sample position with an accuracy better than 1 \um, which was much smaller than the spot size of the incident laser. We illustrate the performance of the sample-position-autocorrection system with representative photoemission data on the topological insulator Bi$_2$Se$_3$ and an optimally-doped cuprate superconductor \Bi. Our method provides new possibilities for studying the temperature-dependent electronic structures in quantum materials by laser-based or spatially resolved photoemission systems with high precision and efficiency.
\end{abstract}

\maketitle

\section{Introduction}
In the past three decades, angle-resolved photoemission spectroscopy (ARPES) has been a leading experimental tool for probing the electronic structure with energy and momentum resolutions in solid materials\cite{Sobota2021}. ARPES systems equipped with laser light sources provide opportunities to realize energy resolution below 1 meV or temporal resolution\cite{Zhou2018}.
It has been revealed that the contact voltage, which results from the different work function between the measured sample and the sample puck, can deflect the photoelectrons and distort the electron trajectories\cite{Fero2014}. Such an effect in the photoemission spectra is evident for low-energy photoemitted electrons with the laser light sources, and thus, a small displacement between the sample and the probe beam would induce noticeable distortions in the photoemission spectra.
In addition, there are usually different terminations on the surface of the sample, such as the topological insulators MnBi$_2$Te$_{4}$/(Bi$_2$Se$_3$)$_n$ (n = 0,1,2...)\cite{Zhong2021a, Wu2020, Hu2020, Vidal2021, Hu2020a} and the high-temperature superconductor \YBCO\cite{Iwasawa2019, Kondo2007}.Therefore, there is growing interest in ARPES experiments with spatial resolution.
Moreover, the spatial resolution is also important in the study of two-dimensional systems, such as the monolayer samples and heterostructures, which are usually obtained by exfoliating bulk materials or by stacking different monolayer samples\cite{Cao2018, Wang2012, Novoselov2016a}. The sizes of the different terminations and artificial two-dimensional samples are typically tens of microns or even smaller.
For the above reasons, it is important to fix the probe beam spot on the sample with a precision much smaller than the beam and sample sizes, which are typically > 1 \um~in regular ARPES experiments.

It is also important to measure the temperature-dependent electronic structures in those materials, which have rich phase diagrams and undergo phase transitions when the temperature is varied, e.g., superconductors. However, in an ARPES system, the thermal expansion and contraction of the cryogenic sample manipulator in the heating and cooling processes can be much larger than the beam size or even the sample itself. The variation of the sample position at temperature between 4 K and room temperature is on the scale of mm, taking a cryostat with the length of 1 m for an example. Specially designed cryostats with flexible connections between the sample and the cold finger are widely used in current  ARPES systems to reduce the thermal drift of the sample in the process of varying the sample temperature. However, the thermal drift of the sample is still much larger than 1 \um, inducing noticeable distortion in the photoemission spectra especially for low-energy probe photons, such as the laser, and making it hard to perform temperature-dependent experiments on tiny samples with sizes on the scale of \um~or tens of \um. Moreover, the design of the flexible connections between the sample and the cold finger comes at the cost of cooling power. Therefore, searching for new methods to fix beam spots on the sample with a precision better than 1 \um~is very important in temperature-dependent ARPES experiments.

In this paper, we introduce a sample-position-autocorrection system with a precision better than 1 \um~on the laser-based ARPES system at Shanghai Jiao Tong University\cite{Yang2019}. Based on the binocular vision method with two cameras mounted on the ultrahigh vacuum chamber and a developed image pattern matching procedure combined with a high-precision four-axis sample manipulator, we managed to track the sample position in the temperature-dependent ARPES experiments with an accuracy better than 1 \um, which is much smaller than the spot size of the incident laser beam. Using flowing liquid helium and equipping a built-in resistive heater, the sample temperature could be varied between 2.5 and 500 K continuously. The performance of the developed sample-position-autocorrection system was benchmarked by measuring the fine temperature-dependent electronic structures on the topological insulator Bi$_2$Se$_3$ and an optimally-doped cuprate superconductor \Bi~(Bi2212).

\section{Photoemission on displaced sample}

\begin{figure}
\centering\includegraphics[width = 1\columnwidth] {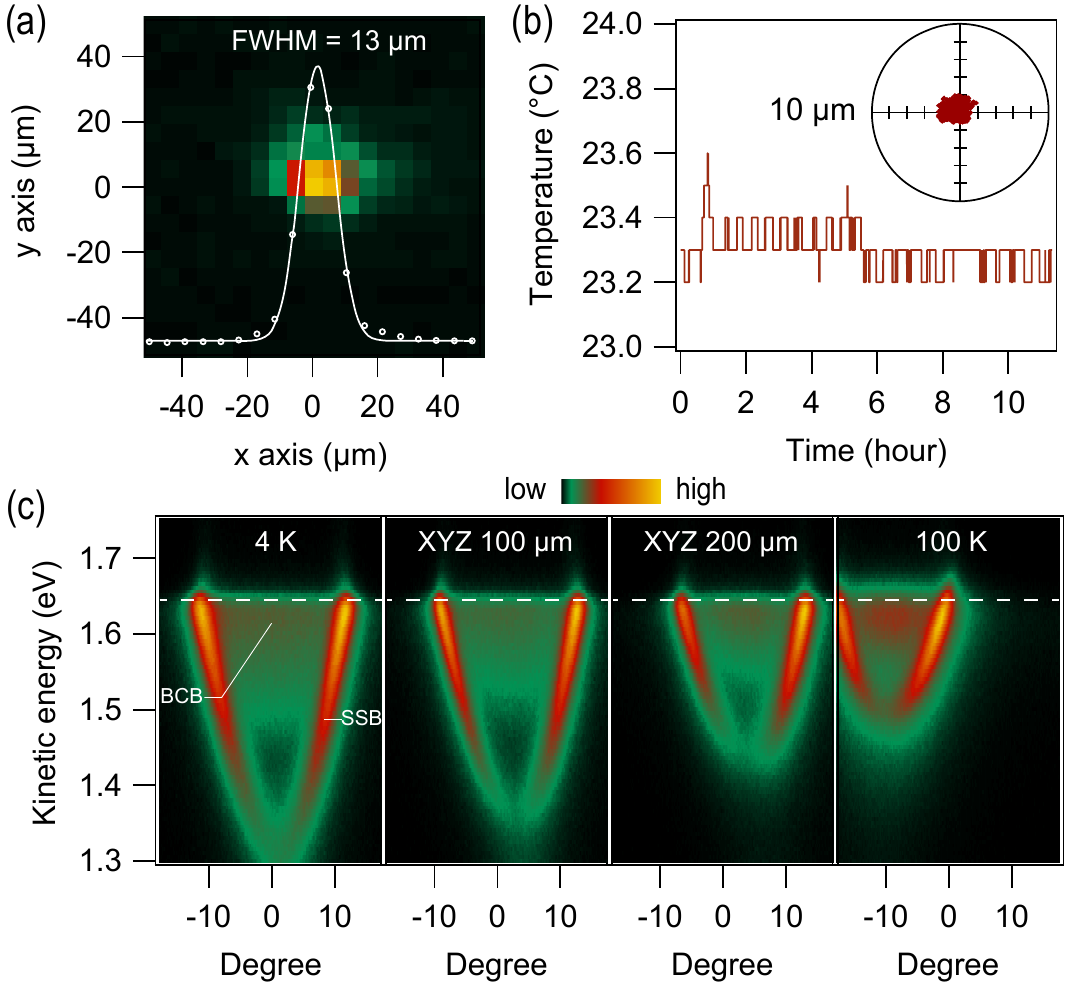}
\caption{
ARPES measurements with the probe beam spot displaced on the sample surface.
(a) Typical beam profile of the ultraviolet laser (6.05 eV, 113 fs) and its line profile along the x-axis sliced at y = 0. (b) Room temperature stability test of the laboratory for 12 h. The inset shows the probe beam pointing test at the sample position. (c) Photoemission intensity as a function of the kinetic energy and the emission angle of a \BS~sample after optimizing the focus of the electron analyzer and the sample position at 4 K after moving the sample along the X-, Y-, and Z-axes for 100 and 200 \um, respectively, and after heating the sample to 100 K without optimizing its position. The white dashed lines indicate the Fermi level.}
\label{Fig1}
\end{figure}

Figure \ref{Fig1}(a) shows the beam profile of the ultrafast ultraviolet laser (6.05 eV) at the sample position in our ARPES system\cite{Yang2019}, on which we recently realized a time resolution of 113 fs, an energy resolution of 16.3 meV, and an ultrahigh energy resolution of 0.4 meV with a 7 eV laser. The full width at half maximum (FWHM) of the beam spot was evaluated by fitting the intensity distribution curve along the x-axis sliced at y = 0 with a Gaussian function (Fig. \ref{Fig1}(a)). The extracted FWHM was about 13 \um, enabling a spatial resolution of about 10 \um~on this system. The probe beam pointing and the flux during long-time experiments were maintained by keeping the room temperature stabilized to within ±0.1 \textcelsius~(Fig. \ref{Fig1}(b)). The beam pointing on the sample was stabilized to within ±1.5 \um~(inset of Fig. \ref{Fig1}(b)), which was limited by the resolution of the used beam profiler, and the flux of the 6.05 eV probe was stabilized to within ± 1.5\% after more than 14 h, which is not shown here.

However, despite the ultrahigh stability of the optical system, the relative motion of the sample away from the beam spot had a tremendous impact on the photoemission spectra. Figure \ref{Fig1}(c) shows the electronic structure of the \BS~measured by the 6.05 eV ultrafast laser at the optimal position (near the center of the sample) and at the temperature of 4 K, where the surface state band (SSB) and bulk conduction band (BCB) are clearly identified. After intentionally moving the sample away from its optimal position along all three translation axes for about 100 and 200 \um, the photoemission spectra showed significant distortions that could be attributed to the uneven surface of the sample or the electric field at the sample edge induced by the different work functions between the sample and the mounting copper base\cite{Fero2014}. Similar distortion was also observed after heating the sample to 100 K, and the much more distorted photoemission spectra indicated that the relative displacement between the sample and probe beam was larger than 200 \um. A sample displacement on the scale of 1 mm was estimated by tracking the sample position with a charge-coupled device (CCD) camera. The distortion was worse at higher temperatures, and there was even no photoemission spectral signal above 200 K because the beam was off the sample. We note that we used a four-axis sample manipulator without flexible connections between the sample and the cold finger, and the estimated expansion of a cryostat with a length of 1.4 m is about 2 mm, which is larger than the size of the sample used.

The above tests showed that the position drift of the sample made it hard to perform precise temperate-dependent ARPES experiments especially for the laser-based ARPES.

\begin{figure*}
\centering\includegraphics[width = 2\columnwidth] {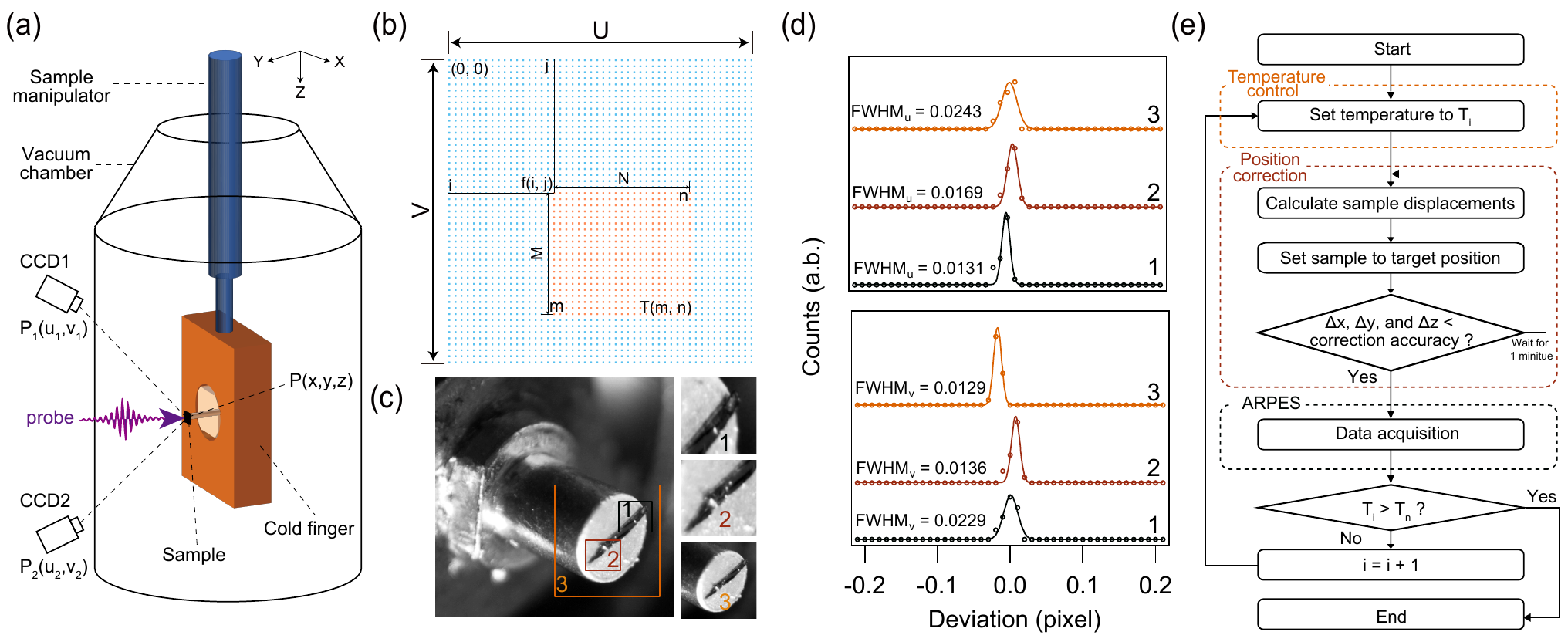}
\caption{
Method of sample position autocorrection in temperature-dependent ARPES experiments. (a) Geometry of the binocular stereo vision system combined with a sample manipulator mounted in an ultrahigh vacuum chamber. (b) Schematic diagram of the grayscale matching calculations. The cyan squares indicate the target range captured by a camera with a size of V$\times$U pixels, and the orange squares indicate the sampling range with a size of M$\times$N pixels. The brightness values of the target image and sampling image are represented by $f(i,j)$ and $T(m,n)$, respectively. (c) Examples of the sample image in an ultrahigh vacuum chamber captured by a camera (left) and selected sampling images (right) for the grayscale matching calculation. (d) Histograms to show the performance of the position autocorrection system based on the real sample image in (c). The target image is shifted randomly (along $u$ and $v$) for thousands of times by interpolating the intensity at each pixel, and the times of each deviation with a step of 0.01 pixel from the real position after correction are counted. Simulations are made with the sampling images 1, 2, and 3 in (c). (e) Flowchart of the temperature-dependent ARPES experiments.}
\label{Fig2}
\end{figure*}

\section{vision tracking method}
To address the issue of thermal drift of the sample position in the temperature-dependent ARPES experiments, we developed a sample-position-autocorrection system based on the binocular vision technique, which included two high-speed industrial cameras with the resolution of 2592 $\times$ 1944, a sensor size of 1/2.5 in, and the pixel size of 3.2 \um, as schematized in Fig. \ref{Fig2}(a). The sample was mounted on the cold finger of the cryostat directly in an ultrahigh vacuum chamber with a pressure better than 4 $\times $ 10$^{-11}$ torr. The sample temperature could be continuously tuned between 2.5 K and 500 K by using flowing liquid helium combined with a built-in resistive heater. The sample position was manipulated by a specially designed four-axis manipulator with three high-precision translation stages and one rotation stage, and all the motions were driven by the stepper motors. The resolution of the translation stages was 1.5 \um, with the minimum achievable incremental movement of 0.5 \um, and the rotation was realized by a rotary seal mounted on top of the translation stage to manipulate the polar angle of the sample. The sample puck with a diameter of about 12 mm was directly screwed into a M8-tapped hole on the cold ﬁnger. There was no flexible connection between the sample and the mounting flange of the cryostat. Therefore, thermal-induced sample translations were the primary problem of our ARPES system. A unified procedure was programmed to control the stepper motors, read the cameras, tune the sample temperature, and collect the photoemission data for our ARPES system. 

In the sample-position-autocorrection system, the sample coordinates in real space are expressed as $P(x, y, z$), and those in the images captured by the two cameras are expressed as $P_{1}$ ($u_1, v_1$), and $P_{2}$ ($u_2, v_2$). The captured image sharpness by the camera was determined based on multiple factors, such as the lighting conditions, resolution of the camera, and camera lens. The incident angle and the brightness of the lights should be tuned to increase the contrast of the detected object and to decrease the oversaturation of the detected region and the reﬂection from the sample surface. The field of view ($FOV$) of the camera is given by $FOV = WD \times L/f$, in which the work distance $WD$ was about 250 mm, $L$ was the sensor size along the horizontal or vertical direction, and the focal length of the imaging lens $f$ was 55 mm. The obtained $FOV$ along the horizontal and vertical directions were about 37 and 28 mm, respectively. The relative motions of the sample in real space and the captured images on the two cameras satisfied
\begin{equation}
	\begin{cases}A_{1}\Delta x+B_{1}\Delta y+ C_{1}\Delta z=&\Delta u_{1}
	\\A_{2}\Delta x+B_{2}\Delta y+C_{2}\Delta z=&\Delta v_{1}
	\\A_{3}\Delta x+B_{3}\Delta y+C_{3}\Delta z=&\Delta u_{2} 
	\\A_{4}\Delta x+B_{4}\Delta y+C_{4}\Delta z=&\Delta v_{2},
\end{cases} 
\label{eq1}
\end{equation}
where the $\Delta x$, $\Delta y$, and $\Delta z$ represent the sample displacements along the X-, Y-, and Z-axes in real space;  $\Delta u_{1}$ and $\Delta u_{2}$ in Eq. \ref{eq1} are the motions in pixels along the horizontal direction in the images captured by the two cameras; and $\Delta v_{1}$ and $\Delta v_{2}$ are the motions in pixels along the vertical directions in the images captured by the two cameras. Thus, once the sample motions on the two cameras ($\Delta u_{1}$, $\Delta v_{1}$, $\Delta u_{2}$, and $\Delta v_{2}$) are determined, the $\Delta x$, $\Delta y$, and $\Delta z$ can be solved from any three equations listed above and the last equation can be used to double check the solutions.
All the coefficients could be determined experimentally in the initial calibration. The remaining question was how to determine the $\Delta u_{1}$, $\Delta v_{1}$, $\Delta u_{2}$, and $\Delta v_{2}$ in the camera images effectively and precisely.

To recognize the sample motions on the images from the two cameras, the image pattern matching method was adopted to track the sample motion. Image pattern matching is a process of finding the same or similar regions in the target image from the template image by analyzing the similarity and consistency of the grayscale, edge, or shape structure. Initially, feature information of the template image was calculated and stored for further image pattern matching calculations. Then, the motion of the template image could be determined after the image pattern matching calculation. 
The algorithm of image pattern matching we employed was the grayscale matching, as illustrated in Fig .\ref{Fig2}(b). During the pattern matching calculation, the template image (M$\times$N) moved from left to right and top to bottom in the target image (V$\times$U), and the matching area between the template image and the target image was determined by the regular cross-correlation\cite{Yamane2020} 

\begin{equation}
	Rcc(i,j) = \frac{\sum\limits^{M-1}_{m=0} \sum\limits^{N-1}_{n=0}f(i+m, j+n)T(m,n) }
	{\sqrt{\sum\limits^{M-1}_{m=0} \sum\limits^{N-1}_{n=0}f^2(i+m, j+n)\sum\limits^{M-1}_{m=0}\sum\limits^{N-1}_{n=0}T^2(m, n) } } ,
\label{eq2}
\end{equation} 
where $f(i,j)$ is the brightness of the target image in the $i^{th}$ row and $j^{th}$ column, and $T(m,n)$ is the brightness of the template image in the $m^{th}$ row and $n^{th}$ column (Fig. \ref{Fig2}(b)). Ideally, $Rcc = 1$, which suggests that the template image is identical to the target image. We note that in the process of image pattern matching, both of the target image and the reference images were interpolated to reduce the impact from different image boundaries. Finally, the sample motions on the two cameras could be precisely determined, and the motion of sample displacements in real space could be calculated according to the coordinate changes on the two cameras by using Eq. \ref{eq1}. Sample displacements were corrected by moving the sample in the direction opposite to the calculated displacements.

To obtain the recognition accuracy of the image pattern matching method based on the grayscale matching from the current setup, we simulated a series of moved images to match with the template image. Figure \ref{Fig2}(c) shows the optical image of the sample at the optimal position in the vacuum chamber. The simulated target image was obtained by moving the optical image (Fig. \ref{Fig2}(c), left) along the horizontal and vertical directions by 0.1 pixels each time. Three square areas (Fig. \ref{Fig2}(c), right) were selected as the template image to perform the image pattern matching calculation. The histograms of the differences between the displacements of the simulated image obtained by the image pattern matching method and the theoretical  movements along horizontal (top) and vertical (bottom) directions for template images 1, 2, and 3 are shown in Fig. \ref{Fig2}(d). The accuracy of the image pattern matching $Res$ is defined as
\begin{equation}
	Res = |peak| + FWHM,
\end{equation}
in which the $peak$ is the position of the maximum counts, and $FWHM$ is the full width at half maximum of the histogram. The $peak$ and $FWHM$ values were  obtained by fitting the histograms in Fig. \ref{Fig2}(d) to Gaussian functions. The estimated accuracies along the horizontal direction for template images 1, 2, and 3 were 0.0153, 0.0236, and 0.0270 pixels, and those along the vertical direction were 0.0238, 0.0211, and 0.0302 pixels, respectively. The simulations showed that the accuracy of the image pattern matching was weakly dependent on the selected template images. In principle, the more feature information the template image contains, the more accurate the matching result will be. In our sample-position-autocorrection system, the pixel sizes of the cameras were 3.2 \um. Thus, 0.0302 pixel was equal to 0.097 \um~in the case of 1:1 imaging, which is much smaller than the beam size of the incident light sources.
	
We next discuss the flowchart of the auto-temperature-dependent ARPES measurements based on the above binocular vision technique.
Figure \ref{Fig2}(e) summarizes the flowcharts of the auto-temperature-dependent electronic structure measurements, which included three steps of (I) sample temperature control, (II) position correction, and (III) ARPES data acquisition. All these procedures were executed step by step, and the lighting illumination was only turned on (controlled by the computer) during the image capture to avoid the illumination-induced thermal effects. Because the sample position could be stabilized in a small amount of time in the temperature-dependent measurements (usually within 1 min), the sample-position-autocorrection program was executed before acquiring the ARPES data. The typical processing time of each position correction was about 2 s, which was dependent on the efficiency of the image pattern matching calculation and the velocity of the stepper motors. The measurements started from an initial temperature $T_0$, and a template image was selected at $T_0$ for the image pattern matching calculation. At the target temperature $T_i$, the sample position was persistently autocorrected by performing image pattern matching calculations and solving for $\Delta x$, $\Delta y$, and $\Delta z$. When the solved $\Delta x$, $\Delta y$, and $\Delta z$ were smaller than a set value of the correction accuracy, taking 1 \um~for an example, the procedure proceeded to perform the ARPES data acquisition. The above process was performed until the sample temperature $T_i$ reached the target temperature $T_n$.
The above temperature control, stepper motor control, image pattern matching calculation, and ARPES data acquisition were programmed and combined into one package of procedures.

\section{Performance tests}

\begin{figure}
\centering\includegraphics[width = 1\columnwidth] {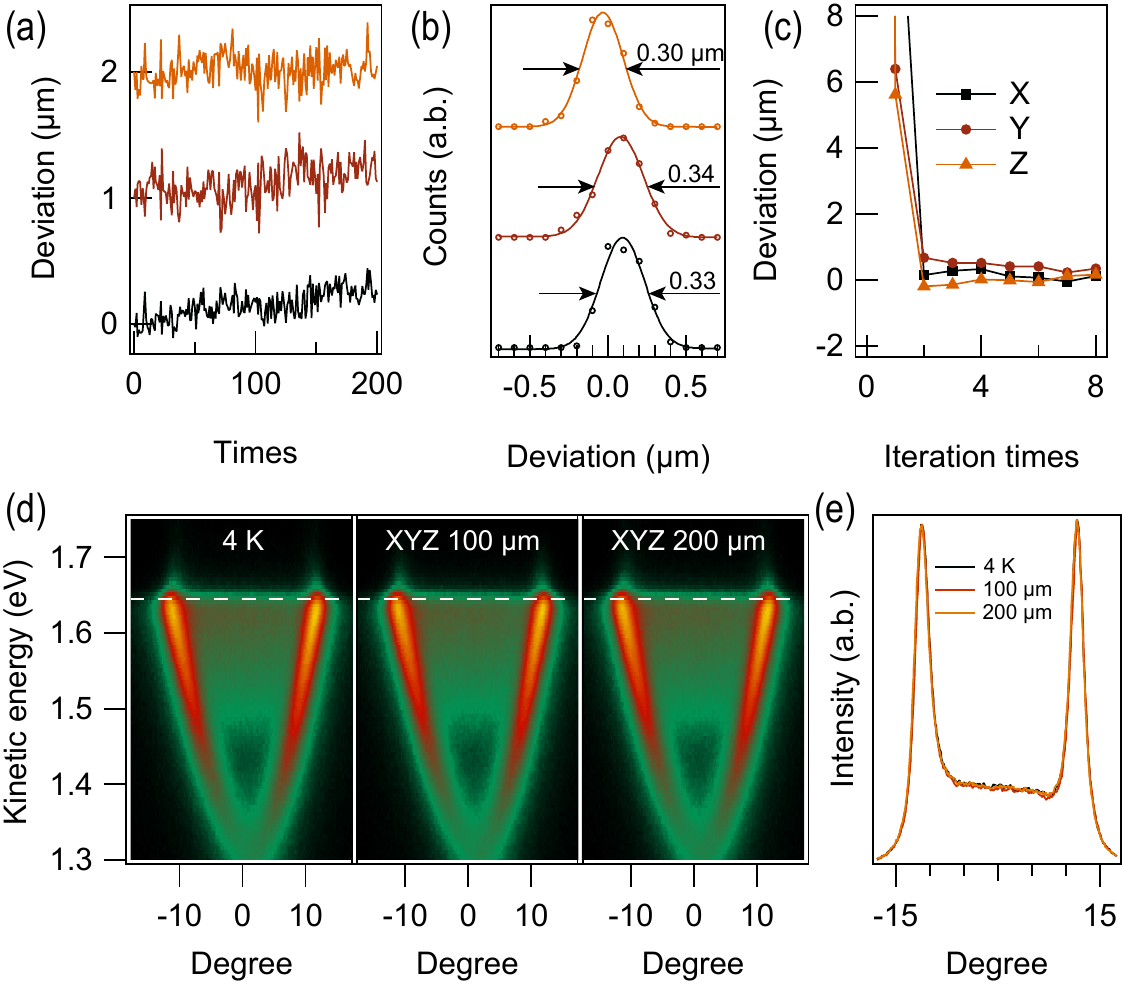}
\caption{
Performance test of the sample position autocorrection system at a fixed sample temperature. (a) Calculated sample position deviations obtained by continually keeping reading the sample position along the X-, Y-, and Z-axes. (b) Histograms of the deviations from (a) for the X-, Y-, and Z-axes. (c) Deviations as a function of the correction time. Offsets were applied on traces in (a) and (b). (d) Photoemission spectra after the sample position was autocorrected. The sample was manually moved by 100 and 200 \um~along all three axes. (e) Photoemission intensity as a function of the emission angle at the Fermi energy from (d).}
\label{Fig3}
\end{figure}

A basic experiment was carried out to investigate the performance of the sample-position-autocorrection system by repeatedly tracking the sample position without any real motion (Fig. \ref{Fig3}(a)). As shown by the histograms (200 times tracking in total) in Fig. \ref{Fig3}(b), the obtained accuracy of the vision tracking system along X-, Y-, and Z-axes were 0.424, 0.425, and 0.337 \um, which were obtained by fitting the histograms with Gaussian functions.
The overall broadening was not simply determined by the accuracy of the image pattern matching process but was also dependent on the stability of the sample manipulator.

Iterations were performed during the image pattern matching process to minimize the displacement in actual experiments.
Figure \ref{Fig3}(c) shows the sample deviations obtained by iterating through the sample-position-autocorrection procedure. Initially, the sample was intentionally moved away from its optimal position along all the X-, Y-, and Z-axes by 0.2 mm. After several iterations, the sample went back to its initial position with a displacement smaller than 1 \um, which was better than the resolution of the translation motors.
Using such a sample-position-autocorrection system, photoemission experiments showed almost identical spectra after executing the sample-position-autocorrection procedure even for an intended displacement of 200 \um~along all the X-, Y-, and Z-axes (Fig. \ref{Fig3}(d)), which is further evidenced in the momentum distribution curves (MDCs) shown in Fig. \ref{Fig3}(e).

\begin{figure}
\centering\includegraphics[width =1\columnwidth] {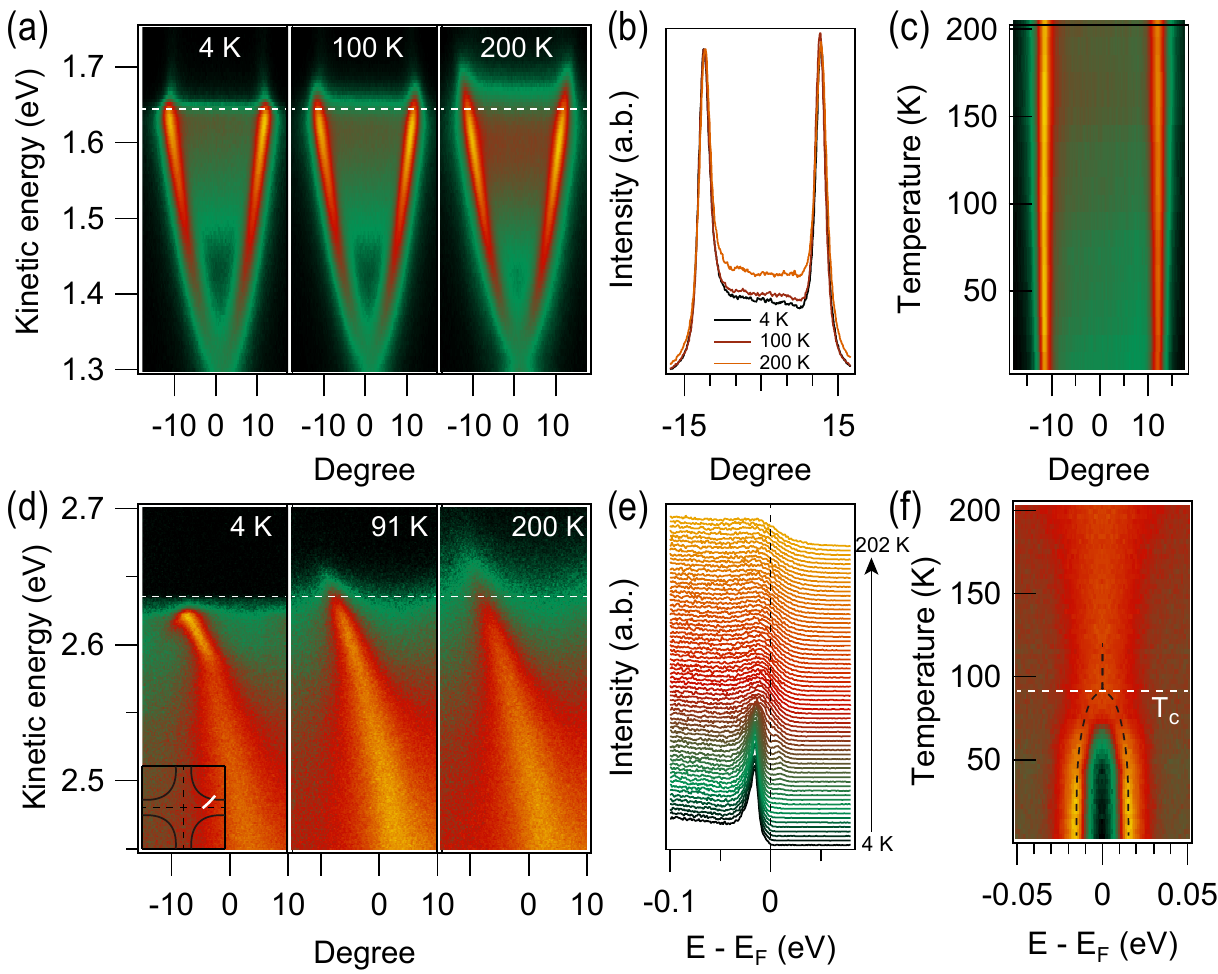}
\caption{
Temperature-dependent photoemission spectra of the \BS~and Bi2212 obtained by using the sample position autocorrection system. (a) Photoemission spectra obtained by heating the \BS~from 4 K to 100 K and 200 K. (b) Photoemission intensity as a function of the emission angle at the Fermi energy from (a). (c) Photoemission intensity as a function of the photoemission angle and temperature. The step of the temperature was 10 K. (d) Similar to (a) with measurements on Bi2212. The inset shows the measured cut (white line) in the Brillouin zone. The white dashed lines in (a) and (d) indicate the Fermi level. (e) Energy distribution curves at the Fermi momentum from 4 to 202 K with a step of 3 K. The intensity was not normalized. (f) Image plots of symmetrized EDCs at the Fermi momentum from 4 to 202 K. The white dashed line indicates the temperature of 91 K, which is the superconducting transition temperature of the Bi2212 sample, and the black dashed lines denote the peak positions.}
\label{Fig4}
\end{figure}

Detailed temperature-dependent ARPES measurements were carried out on the topological insulator \BS~to show the performance of the sample-position-autocorrection system. The measurement on the \BS~was performed using the 6.05 eV ultrafast laser, and during the measurement, the sample temperature was varied from 10 K to 200 K with a step of 10 K. As the temperature increased, no visible distortion appeared in the photoemission spectra (Fig. \ref{Fig4}(a)), as further evidenced by the MDCs at the Fermi energy in Fig. \ref{Fig4}(b) and also by the intensity as a function of the temperature and photoemission angle in Fig. \ref{Fig4}(c). Enhanced photoemission signals were evidenced from the bulk band at high temperatures near the Brillouin zone center.

High-precision temperature-dependent electronic structure measurements are important in studying those quantum materials with electronic phase transitions, taking the superconductor as an example.
In Figs. \ref{Fig4}(d--f), temperature-dependent measurements were taken on the optimally-doped high-temperature superconductor Bi2212 with the 7 eV ultrahigh resolution laser, which provided an overall energy resolution of 0.4 meV. By taking the advantage of the sample-position-autocorrection system, no visible shift of the Fermi momentum due to the curvature of the cleaved surface or the electrical field around the sample was evidenced from 4 to 200 K (Fig. \ref{Fig4}(d)), making it convenient to study the photoemission spectrum as a function of temperature at the Fermi momentum. The energy distribution curve (EDC) at a fixed photoemission angle (the Fermi momentum) showed a clear energy gap feature and a Bogoliubov quasiparticle peak\cite{Kondo2015, Zhang2012a, Sun2018a} above the Fermi energy below the superconducting transition temperature (Fig. \ref{Fig4}(e)). The symmetrized EDC as a function of the temperature clearly showed an energy gap below the transition temperature and the gap gradually increased upon cooling the sample until reaching a constant value at very low temperatures (Fig. \ref{Fig4}(f)).

The above performance tests clearly showed the successful application of the sample-position-autocorrection system in the temperature-dependent ARPES experiments. We have applied such high-precision temperature-dependent ARPES experiments on the iron-based superconductor FeSe$_{1-x}$S$_x$, in which there are mutually perpendicular orthorhombic domains below the structural transition temperature. The sample-position-autocorrection system enabled us to perform photoemission tests in a specific domain over a large temperature range to reveal the electronic phase transitions\cite{Yang2022a, Yang2022b}.  The developed sample-position-autocorrection system will be potentially useful for spatially-resolved ARPES, enabling temperature-dependent electronic structure measurements in various materials, including micro-sized 2D materials obtained by mechanical exfoliation, artificial heterostructure, materials with multiple domains, and so on. We hope that the performance of such a position-autocorrection system in real experiments can be benchmarked by the ARPES experiment with spatial resolution in the future \cite{Kitamura2022, Iwasawa2017, Mao2021}.

\section{Conclusion}
In  this study, we have developed a high-precision sample-position-autocorrection system for use in ARPES experiments. Based on the binocular vision technique, we were able to track the sample position in real time in ARPES experiments with an accuracy better than 1 \um, which was much smaller than the beam spot of the current incident light. The successful application of such a system made it possible to take high-precision temperature-dependent electronic structure measurements, which were of great significance for studying the electronic phase transitions in quantum materials. This sample-position-autocorrection system was also applicable for other experiments in which the sample position needed to be controlled precisely, such as  the optical reflectivity experiment, the electron diffraction experiment, and so on. 
The position tracking accuracy of the current study was limited by the resolution of the sample manipulator and the vision camera. As such, it would be interesting to reveal the physical limit of the accuracy based on such a binocular vision technique in ARPES experiments. Moreover, the presented sample-position-autocorrection system was simplified, and only 2D image pattern matching  was adopted by using a four-axis sample manipulator. Further study with the development of a possible state-of-art 3D scene reconstruction from 2D images could be useful for correcting the full six degrees of freedom of the sample.

\begin{acknowledgements}
We thank X. J. Zhou, Y. H. Wang, and D. Qian for providing Bi$_2$Se$_3$ samples, Y. Huang, J. S. Wen and G. D. Gu for providing \Bi~samples. W. T. Z. acknowledges support from the National Key R\&D Program of China (Grants No. 2021YFA1401800 and No. 2021YFA1400202) and National Natural Science Foundation of China (Grants No. 11974243 and No. 12141404) and Natural Science Foundation of Shanghai (22ZR1479700) and additional support from a Shanghai talent program.
\end{acknowledgements}

\end{document}